\documentclass{article}
\usepackage{arxiv}
\usepackage[utf8]{inputenc}

\usepackage{tikz}

\usepackage{hyperref}
\usetikzlibrary{matrix}

\usepackage{stfloats}
\usepackage{colortbl}

\def\SC{{\bf SC}}
\def\VACC{{\bf VaccSC}}

\title{Towards reliable and transparent vaccine phase III trials with smart contracts}

\def\UFU{Federal University of Uberlandia}
\def\FACOM{School of Computer Science}

\author{
  Ivan da Silva Sendin \\
  \FACOM\\
  \UFU\\
  \texttt{sendin@ufu.br} \\
   \And
 Rodrigo Sanches Miani \\
  \FACOM\\
  \UFU\\
  \texttt{miani@ufu.br} \\

}

\begin{document}

\maketitle

\begin{abstract}
    Transforming a vaccine concept into a real vaccine product is a complicated process and includes finding suitable antigens and regulatory, technical, and manufacturing obstacles. A relevant issue within this scope is the clinical trial process. Monitoring and ensuring the integrity of trial data using the traditional system is not always feasible. The search for a vaccine against the coronavirus SARS-CoV-2 illustrates this situation. The scientific credibility of findings from several vaccines' clinical trials contributed to distorted perceptions concerning the benefits and risks of the drug. This scenario is ideal for applying technologies such as Blockchain and Smart Contracts in healthcare issues. This paper proposes a protocol based on Smart Contracts, named VaccSC, to enable transparency, accounting, and confidentiality to Phase III of vaccine experiments. The protocol was implemented in Solidity language, and results show that the VaccSC enables double-blindness, randomization, and the auditability of clinical data, even in the presence of dishonest participants.
\end{abstract}

\section{Introduction}

The COVID-19 pandemic that hit the world in 2019/2020 \cite{Dong2020} is undoubtedly a disruptive event. Basically, all aspects of our society have been affected - and still will be - due to this pandemic. During the COVID-19 pandemic, vaccine development underwent irreversible changes. Before 2019, it was expected that a vaccine would take a few years or even a decade to develop and be considered safe, efficient, and made available to the market \cite{buckland2005process}. Contrary to initial expectations, several laboratories \cite{Doshi2020} released advanced studies of COVID-19 vaccine candidates in less than a year of research. This speed may raise doubts about the honesty and rigor of these scientific experiments.

Before regulatory approval, a vaccine candidate usually experiences three phases of development in humans: Phase I, Phase II, and Phase III. During Phase III large-scale trials, the human subjects are divided into two groups: i) those who received the vaccine and ii) those who received a control, a placebo, for example. The main outcome of this phase is the vaccine efficiency with is calculated by comparing the incidence of disease in vaccinated subjects and the incidence of disease in unvaccinated subjects (control groups). Unfortunately, this process is not transparent to society as a whole, which can compromise acceptance of the vaccine by some individuals \cite{Mahase2020,Harrison2020}.

While transparency is important, the confidentiality of emerging data is also relevant: vaccine developers should not have access to them with the risk of changes in the trial design and compromising the final research results \cite{Krause2020}. The recent technologies of Blockchain and Smart Contracts are primarily known for adding a layer of "trust" on the Internet services, as they offer decentralization, immutability, and public access to stored data \cite{swan2015blockchain,Xiwei}.

The use of Blockchain and Smart Contracts in the health area is not new. Several studies point out the possibility of using these technologies in healthcare sustems\cite{Zou2020,Reen2019}, diagnostics tracking\cite{Choudhury2020}, vaccine and medication trial tracking\cite{wong2019prototype,benchoufi2017blockchain}. More recently, the so-called "COVID-19 passports" were addressed by these technologies \cite{Eisenstadt2020}, \cite{Halpin2020} \cite{Shamsi2020}. The previously cited approaches maintain trust "out of the contract". In other words, the data is stored in the contract, and participants use it with a notary system. Therefore, the contract cannot interfere with the integrity of the stored information.

In this work we present a protocol based on smart contracts and Commitment Schemas to enable transparency, accounting and confidentiality to Phase III of vaccine experiments. Our work's main contribution for the use of Smart Contracts in vaccine trials is the use of Commitment Schemas to enable the properties of double-blind, randomization and the auditability of clinical data, even in the presence of dishonest participants. Also, the Smart Contracts can determine whether the experiment has reached the desired efficiency and decide on the vaccine approval.

The rest of paper is organized as follows. In Section 2 we outline the theoretical background of our proposal which includes a description of Commitment Schemes and Smart Contracts. We briefly describe the vaccine development process in Section 3. Section 4 details the proposed protocol called \VACC. Section 5 summarizes related work and Section 9 concludes the paper.

\section{Background}

\subsection{Commitment Schemes and Coin Flipping}\label{sec:commit}

Commitment Schemas emerged in the early 1980s \cite{Blum1983} and are used when a party  $\mathcal{A}$  needs to commit to a certain amount $v$ with another party $\mathcal{B}$  without revealing the value of $v$ immediately.

A common construction of Commitment Schemes is done using cryptographic hashing functions and has two phases:

\begin{description}
\item [Commit Phase] In this phase, $\mathcal{A}$ commits to the value $v$ generating a nonce\footnote{Nonce is a pseudo-random number used only once in some cryptographics protocol.} $n$ and producing $$\mathrm {C} = \mathrm{H}(v|n),$$ where $ | $ indicates the concatenation operation and $\mathrm{H}$ is a cryptographic hash function. The value $\mathrm{C}$ - the  commit - is then made public;

\item [Reveal Phase] To reveal the committed amount, $\mathcal{A}$ sends to $\mathcal{B}$ $v$ and $n$, which calculates the hash value of the received data, and checks whether the generated hash is equal to $\mathrm{C}$.

\end{description}

We say that $\mathrm{C}$ has the properties of {\bf bidding}, since $\mathcal{A} $ cannot create $v^\prime$ and $n^\prime$ that produce the same $\mathrm{C}$; and {\bf hidding} because from $\mathrm{C}$ it is not possible to determine $ v $. The value $\mathrm{C}$ can be seen as a {\em sealed} version of the value $n$ in the sense that once $\mathrm{C}$ has been published, $\mathcal{B}$ does not knows $v$ and $\mathcal{A}$ can no longer change $v$ without being discovered. See \cite{Oded} for a review and security recommendations.

Using the Commitment Schemes, one can build a protocol to establish a pseudo-random number between parties that do not trust each other. This protocol is known as "coin flipping over telephone". To generate a pseudo-random bit, parts A and B follow the following steps:
\begin{itemize}
    \item $\mathcal{A}$ (and $\mathcal{B}$), privately, generate a pseudo-random bit $R_\mathcal{A}$($R_\mathcal{B}$);
    \item $\mathcal{A}$($\mathcal{B}$) commits and reveals $R_\mathcal{A}$($R_\mathcal{B}$) using a Commitment Scheme;
    \item Now, $\mathcal{A}$ and $\mathcal{B}$ knows both $R_\mathcal{A}$ and $R_\mathcal{B}$, and both can calculate $R=R_\mathcal{A} \oplus R_\mathcal{B}$;
    \item The resulting bit $R$ is a pseudo-random bit shared by $\mathcal{A}$ and $\mathcal{B}$.
\end{itemize}

One of the parties acting honestly is enough for the bit produced by the protocol to be random\cite{Blum1983}.

\subsection{Smart Contracts}\label{sec:sc}

Introduced by Nick Szabo in \cite{SzaboSC}, Smart Contracts (\SC) are programs whose correct execution cannot be prevented. A \SC~ can be compared to vending machines: with a coin and the press of a button, the user gets the expected product with a certain degree of confidence without dependence on other entities.

Currently, the \SC~ platforms are obtained through P2P networks that execute contracts - also called {\em dapps} - redundantly. The participating nodes in this network receive financial incentives for the correct execution of these contracts. The high degree of redundancy and the economic stimulus produces an extremely reliable virtual computer.

The main \SC~ platform currently available is Ethereum \cite{wood2014ethereum}, which uses its own cryptocurrency, called Ether, in its transactions and to reward participating nodes. Ethereum is permissionless, meaning that anyone can access the network. Users at the Ethereum platform are identified by an address that is essentially a public key. Access to the system, such as transferring values or invoking methods in contracts, is done through signed messages using the private key corresponding to Ethereum addresses. The confidence in the correct execution of programs and the extensive use of cryptographic protocols allow \SC~ to be secure, transparent, and auditable.

The applications of \SC~ are countless: financial services, games, betting, and network infrastructure are some examples\footnote{See \url{https://etherscan.io/dapp} for an extensive list of {\em dapps}.}. The capability to enforce rules, accurately described by a programming language, allows the adoption of \SC~ in other application scenarios such as public administration \cite{Krogsboll2020} and regulated activities \cite{Andoni2019}.

\section{Vaccine development proccess}

The vaccine development goes through an elaborate development process after the discovery stage \cite{singh2016clinical}. It is common to divide this process into preclinical and clinical stages. The clinical-stage is defined by several clinical trials in human subjects organized by three phases. The objective of phase I is to evaluate the vaccine's safety in a small sample of human volunteers. During phase II, the goal is to demonstrate immunogenicity vaccine by assessing the impact of multiple variables on immune response, such as age, ethnicity, gender, and presence of maternal or pre-existing antibodies. The sample of human volunteers in phase II varies from hundreds to thousands of subjects. Phase III trials are large-scale clinical trials and essential for registration and approval to license the vaccine. If the phase III results demonstrate efficacy and safety, the vaccine manufacturer can apply to the national regulatory authority to approve and market the product. The primary outcome is vaccine efficiency (VE), which can be defined as \cite{nauta}:

\begin{equation}
  VE = (1 - RR)*100\%
\end{equation}
where RR represents risk reduction (or relative risk). RR is the ratio between the incidence of disease in vaccinated subjects and disease incidence in unvaccinated subjects. Considering $AR_{1}$ the number of infected cases among the subjects in the investigational vaccine group and $AR_{0}$ the number of infected cases among the control subjects then 
\begin{equation}
RR = \frac{AR_{0} - AR_{1}}{AR_{0}}.
\end{equation}
This can be seen as a comparison of who received the vaccine and who received a placebo.
To illustrate how the RR is calculated we will use the \cite{pfizer}: a threshold of 164 contamination confirmations was determined for this experiment. If, for example, after 164 infections it is found that 120 were from people who received the control and 44 from people who received the vaccine, the RR of this vaccine would be approximately 0.63.

The disease's occurrence is the most common endpoint; however, the trial may be based on other clinical endpoints, such as incidence of infection or immunological correlations of protection. After completing Phase III trials and following licensure of the product, Phase IV studies, also referred to as postmarketing surveillance studies (PMS), are used to monitor the vaccine for safety and population effectiveness.

The well-accepted standard for carrying out vaccination experiments is conducted in the randomized controlled, double-blind trial system \cite{rct}. In this system, the person responsible for applying the vaccine randomly chooses what will be applied to each individual, not knowing its content (vaccine or control). Also, the subject who receives the shot does not know what he/she is receiving. This procedure aims to avoid bias in the vaccination process. Eventually, a dishonest vaccine company would like to improve its product's efficiency by selecting a healthier subpopulation to receive the vaccination and a less healthy population to receive control. Similarly, the person receiving the shot should not know its content to avoid the experiment's placebo effect.

\section{Proposed Protocol}

In this session, we present \VACC: a \SC~ based protocol to track a Phase III vaccine trial providing transparency and double-blind behavior.

For the sake of simplicity, the proposed protocol models the participation of three types of entities:

\begin{description}
    \item [Vaccine Developer] Responsible for distributing vaccines to clinics and protecting information about their content. He is also responsible for deploying the contract, being the only participant who can distinguish the real vaccine from the control, creates the commits for the shots, which will be used as identifiers in the contract;
    \item [Vaccine Clinic] Apply the shots and updates the \VACC~ biding the shot with a specific patient;
    \item [Patient] Receives the shot and notifies the contract when he/she becomes ill.
\end{description}

We assume that all participants have an Ethereum address and keep the corresponding private key securely. This step can be achieved using a smartphone, for example.

\subsection{Random Patient/Shot Assignment}

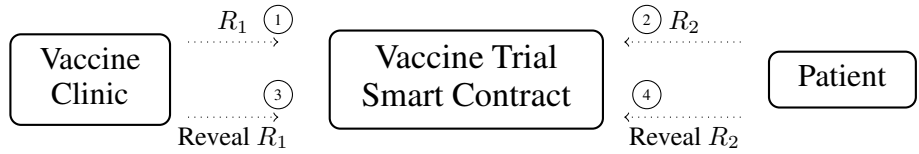
\begin{figure*}[h]
  \centering
  
    \begin{tikzpicture}

\node[draw,thick,rounded corners] at (1.5,2.5) [scale=1.2] {
  \begin{tabular}{c}
    Vaccine  \\
    Clinic
\end{tabular} };

\node[draw,thick,rounded corners] at (11.5,2.5) [scale=1.2] {
  \begin{tabular}{c}
    Patient 
\end{tabular} };

\node[draw,thick,rounded corners] at (6.5,2.5) [scale=1.3] {
  \begin{tabular}{c}
    Vaccine Trial  \\
    Smart Contract
\end{tabular} };

%circulos
\node (s1) [draw,circle,scale=0.6] at (4,3.3) {1};
\node (s2) [draw,circle,scale=0.6] at (8.9,3.3) {2};
\node (s3) [draw,circle,scale=0.6] at (4,2.3) {3};
\node (s4) [draw,circle,scale=0.6] at (8.9,2.3) {4};
%\node (s5) [draw,circle,scale=0.6] at (4,3.3) {5};

\draw[->,dotted] (2.8,3.0) -- (4.0,3.0) node[pos=0.5,above] {$R_1$};
\draw[->,dotted] (2.8,2.0) -- (4.0,2.0) node[pos=0.5,below] {Reveal $R_1$};

\draw[<-,dotted] (8.6,3.0) -- (10.2,3.0) node[pos=0.5,above] {$R_2$};
\draw[<-,dotted] (8.6,2.0) -- (10.2,2.0) node[pos=0.5,below] {Reveal $R_2$};

\end{tikzpicture}

    \caption{Each participant privately generates a random number and commits this number. After both numbers are committed to the contract, each participant reveals the value (steps 3 and 4). The \SC~ combines these numbers - $R_1 \otimes R_2$ - to associate one of the shots available to the patient.}
    \label{fig:random}
\end{figure*}

As stated earlier, the random association between the shot and the patient prevents bias. The generation of random numbers on \SC~ environment is complicated and must be done carefully. In this protocol, we chose to use the approach in which each participant generates a random number ($R_1$ and $R_2$). The SC combines these numbers using bitwise {\bf XOR} operation. With this approach, one part being honest is enough for the number to be random. This process is described in Figure \ref{fig:random}.

\subsection{Double Blindness}

Here we describe how the \SC~ guarantees Double Blindness property. The Vaccine Clinic does not know what he is applying as much as the patient does not know if he receives the vaccine or a control shot.
Initially, Vaccine Developer protects the information on the content of each shot that will be part of Phase III as described in Section \ref{sec:commit}:
$$C_i = \mathrm{H}(n_i|v_i),$$
the value $C_i$  is used as a shot identifier by the contract and must be physically associated with the shot. The values $n_i$ and $v_i$ are saved and used later to reveal the shot's content and calculate the vaccine effectiveness.

Also, when deploying the contract, Vaccine Developer informs:

\begin{description}
    \item [Number of participants:] corresponding to the number of available shots;
    %\item [Threshold of infected:] number of patients who need to be infected in order to calculate VE;
    \item [Infected Threshold:] number of patients who need to be infected before the committed values are revealed and the vaccine's efficiency is calculated; 
    \item [Target efficiency:] minimum VE required in order to vaccine be approved;
    \item [Vaccination Clinics:] the Ethereum address of clinics that will apply the shots and update the \VACC.
\end{description}

The data structure that stores the necessary information to calculate vaccine efficiency are shown in Table \ref{tab:struct}. The fields are presented in chronological order in which they are changed during the execution of the contract.

%\begin{table}[]
% \caption{Data structure used by \VACC~ to mantain vaccine trial information.}
%\label{tab:struct}
%\begin{center}
%\begin{tabular}{|>{\cc}m{75pt}|>{\cc}m{75pt}|>{\cc}m{135pt}|} 

% \hline
%    {\bf Information} & {\bf Type} & {\bf Access and Control} \\
%    \hline
%    Commit & hash & Immutable, crated by Vaccine Developer  \\
%    \hline
%    Clinic & Address   & Each clinic is added by Vaccine Developer \\
%    \hline
%    Patient & Address  & Clinic and patient\\
%    \hline
%    gotSick &  Boolean &  Patient\\
%    \hline
%    Vaccine Type & Vaccine/Placebo & \VACC~ fills field with the  {\em reveal} provided by Vacine Developer  \\
%    \hline

%\end{tabular}\vspace{0,1cm}\\
%\end{center}

%\end{table}

\begin{table}[]
    \centering
    \begin{tabular}{|c|c|l|}
    \hline
    {\bf Information} & {\bf Type} & {\bf Access and Control} \\
    \hline
    Commit & hash & Immutable, crated by Vaccine Developer  \\
    Clinic & Address   & Each clinic is added by Vaccine Developer \\
    Patient & Address  & Clinic and patient\\
    gotSick &  Boolean &  Patient\\
    Vaccine Type & Vaccine/Placebo & \VACC~ fills field with the  {\em reveal} provided by Vacine Developer  \\
    \hline
\end{tabular}

    \caption{Data structure used by \VACC~ to keep vaccine trial information.}
    \label{tab:struct}
\end{table}

\subsection{Execution}

Once the Vaccine Developer deployed the \VACC, the protocol follows the following steps:

\begin{enumerate}
    \item The Vaccine Developer  delivers the shots to each Vaccination Clinic and makes the association between the shots and the clinics in the contract;
    \item The clinic applies the shot and associates the shot identifier with a specific patient. The patient confirms this to contract;
    \item If the patient becomes ill, he communicates the fact to the contract;
    
    \item When the infected threshold is reached, the contract creates an event informing the Vaccine Developer;
    
    \item Using the Ethereum addresses of infected patients, Vaccine Developer selects the corresponding commits. After that, the Vaccine Developer sends information about patients who {\bf received the control} to the contract. The contract calculates the number of patients who received the vaccine;
    
    \item Vaccine efficiency can be determined (using Equation 2), and its approval status can be obtained directly from the contract by anyone.
\end{enumerate}

\begin{figure*}[h]
    \centering
    \begin{tikzpicture}

\node[draw,thick,rounded corners] at (1.5,4.5) [scale=1.2] {
  \begin{tabular}{c}
    Vaccine \\
    Developer
\end{tabular} };

\node[draw,thick,rounded corners] at (6.5,4.5) [scale=1.2] {
  \begin{tabular}{c}
    Vaccine  \\
    Clinic
\end{tabular} };

\node[draw,thick,rounded corners] at (11.5,4.5) [scale=1.2] {
  \begin{tabular}{c}
    Patient 
\end{tabular} };

\node[draw,thick,rounded corners] at (6.5,1.5) [scale=1.5] {
  \begin{tabular}{c}
    Vaccine Trial  \\
    Smart Contract
\end{tabular} };

\node (s1) [draw,circle,scale=0.6] at (3.8,5.5) {1};
\node (s2) [draw,circle,scale=0.6] at (8.8,5.5) {2};
\node (s3) [draw,circle,scale=0.6] at (9.3,3.5) {3};
\node (s4) [draw,rectangle,scale=0.6,thick] at (3,2) {4};
\node (s5) [draw,circle,scale=0.6] at (4,3.3) {5};
\node (s6) [draw,circle,scale=0.6] at (6,0) {6};

\draw[->] (2.8,4.5) -- (5.0,4.5) node[pos=0.5,above] {Vaccine Shot};

%\node (s1) [draw,circle,scale=0.6] at (8.25,5.9) {2};

\draw[->] (7.6,4.5) -- (10,4.5) node[pos=0.5,above] {Vaccinates};

\draw[->,dotted] (11.0,4.0) -- (9,2) node [pos=0.5,right] {Got sick};

\draw[->,dashed,thick] (3.75,1.75) -- (1.5,3.5) node [pos=0.5,left] {Trial Finished};

\draw[<-,dotted] (4,2.5) -- (2.5,3.5) node [pos=0.5,right] {Reveal};

\draw[->,dotted] (6.5,0.5) -- (6.5,-1) node [pos=0.5,right] {Vaccine Status};

%\draw[->,dotted] (4.25,4.75) -- (5,4) node [pos=0.5,above] {Update};

%\draw[->,dotted] (8,4.75) -- (8,4) node [pos=0.5,above] {Update};

\end{tikzpicture}

    \caption{Protocol events associated with one shot. Step four occurs only once. Solid lines indicates physical events whith contract updates, and dashed lines indicates \VACC~ interactions.}
    \label{fig:protocol}
\end{figure*}
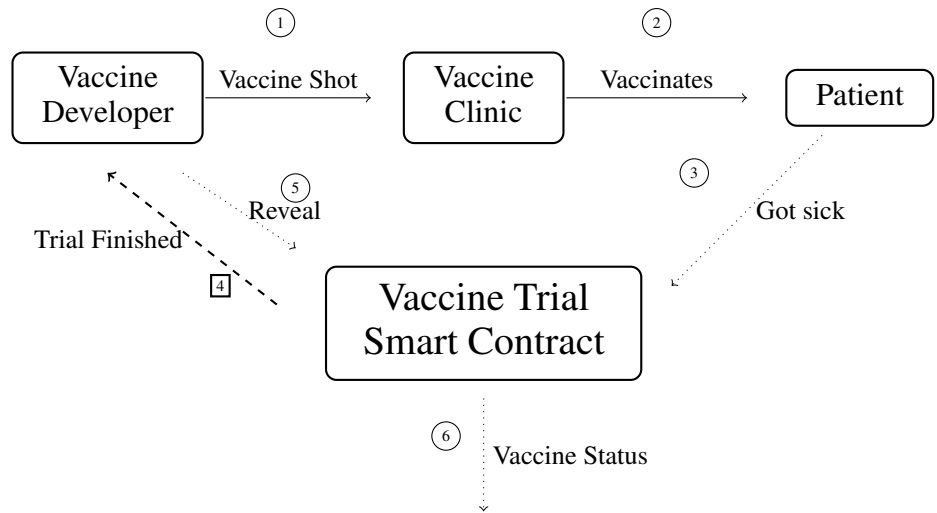

An overview of the protocol is shown in Figure \ref{fig:protocol}.

\subsection{Security analysis}

In the associating vaccines to patients phase, both participants are equally responsible for generating the random number, and one of the participants acting honestly is sufficient to produce a random number. If the two conspire to choose a specific shot, they still cannot choose between vaccine or control, as the Vaccine Developer protected this information.

The double blindness property has the following analysis:

\begin{description}
    \item [Patient] The action expected by the patient is to communicate his own disease event to the contract. Patients do not know about their vaccination status, so they cannot effectively change the results of the contract;
    \item [Vaccine Clinic] As he applies the shot but does not know its content, he is also unable to defraud the protocol;
    \item [Vacccine Developer] A dishonest developer can improve the efficiency of the vaccine by decreasing the number of people who have been effectively vaccinated and have become ill. Using the proposed protocol, he cannot do that because his action will precisely reveal the patients who received a placebo, so the omission is harmful to him. Therefore, the forging of "control patients" does not occur because the shot status is committed.
\end{description}

\subsection{Availability}

The \VACC~ protocol was implemented in Solidity language, the client side applications were developed using the Brownie suite\footnote{\url{https://github.com/eth-brownie}} and Python language.
The contract is available for download at \url{https://github.com/ivansendin/Vaccine}. Commit Scheme and Coin Tossing libraries are available at \url{https://github.com/ivansendin/CommitExamples}.

\section{Related Work}

Early attempts to apply \SC~ to  conduct clinical experiments use \SC~  to store data securely, guaranteeing the data's authenticity and integrity through the access control and digital signatures provided by the platform. There are several works available with this approach \cite{nugent2016improving,benchoufi2017blockchain,omar2019exploiting,wong2019prototype,omar2020ensuring,peng2020efficient}. 

\cite{nugent2016improving} used \SC~ as a solution to data manipulation issues common to clinical trials. The authors propose a hierarchical arrangement of two core types of smart contracts: i) a regulator contract, holding a data structure containing clinical trial authorization (CTA) details, and ii) a trial contract deployed by contract research organizations (CROs). The trial contract's logic effectively enforces aspects of the trial protocol, securing that neither subjects nor measurements are added outside the trial timelines. At the same time, the tamper-resistant characteristics of the blockchain prevent data manipulation. Authors deployed contracts using a private Ethereum blockchain with synthetic data representing the clinical trials of Tamiflu, an influenza drug. They were able to query the state of data, such as the number of trials underway, the number of subjects recruited, and the timestamp at which the transaction was processed.

\cite{benchoufi2017blockchain} attempted to clarify the three main principles of using blockchain in clinical trials: timestamping, time-ordering and smart-contracting. Regarding \SC~, authors focus on providing examples of how it can be applied to clinical trials. According to them, \SC~ can be applied in the entire process. The case of randomization of a patient in a clinical trial before written consent is briefly discussed as a potential application. The paper does not provide any experiments or \SC~ implementations.

Omar et al. \cite{omar2019exploiting} and \cite{omar2020ensuring} proposed a \SC~ using Ethereum to tackle data management challenges in clinical trials. The proposed \SC~ initially captured three stages of a clinical trial process: new drug application, clinical trial initiation, and patient enrollment stages. \cite{omar2020ensuring} discussed patient monitoring and severe adverse event occurrence. The smart contract was written in Solidity and captures interactions among clinical trial stakeholders such as the regulatory agency, drug sponsors, physicians, and patients. For instance, for the new drug application scenario, the smart contract implements the process in which the sponsor obtains permission from the FDA (Federal Drug Agency) to test their drug in humans. The contract has a function that requests the sponsor to upload the required documentation. When the documentation status is set to true, the FDA announces the approval or rejection of the request.  

\cite{wong2019prototype} developed a blockchain-based web portal to tackle challenges in the current clinical trial process. The designed portal is accessible to all interested parties, facilitating the interaction between patient and clinical investigator. The portal also enables expediting adverse event reporting. The authors modeled a phase II clinical trial during their experiments. They simulated how a previously completed clinical trial (omalizumab drug) would behave using blockchain software technologies. Despite arguing that forcing all participating parties to use a service like that is still a challenge, they showed that data entry, storage, and adverse event reporting could be performed more robustly and securely.

\cite{peng2020efficient} addressed the issue of centralized vaccine production supervision. The authors proposed a decentralized blockchain-based method to enforce privacy protection of vaccine production records. The blockchain structure is divided into two levels. The first one is the private data of the pharmaceutical company and consists of production records. The second one is public data. A public blockchain is used to store this data. Every transaction in the block represents a process for each batch of vaccine within the production process. Each transaction also has its timestamp and signature of the company and the hash value of the previous transaction to prevent production data fraud. The authors were able to protect the privacy of the company's production of vaccines using this structure. The proposed framework was implemented using the Hyperledger Fabric blockchain. The authors only evaluated the spatio-temporal efficiency (impact of the transaction against latency, throughput, and blockchain size) of the method.

\section{Conclusion}

We presented \VACC: a protocol based on Smart Contracts that brings transparency and reliability to Phase III of vaccine trials while maintaining the double-blind property present in existing protocols.
The protocol prevents the Vaccine Developer from hiding relevant information to vaccine trials and allows anyone to audit the trial data. The protocol achieves these conflicting requirements using Commitment Schemas.

The presented contract models a simplified version of the real world, other scenarios - such as the presence of observers to ratify the events - can be easily added to the contract. Ultimately, the proposed protocol captures the essence of Smart Contracts - "Code is law" - and can be used as a deciding element on the approval of a vaccine, whether by a country or by an individual.

\bibliographystyle{plain}
\bibliography{mybib}

\end{document}